\documentclass[12pt]{article}

\usepackage{epsfig}
\usepackage{psfrag}
\usepackage{latexsym}
\usepackage[DIV13]{typearea}
\usepackage{amsmath}
\usepackage{amssymb}
\usepackage{amsfonts}
\usepackage{cite}
\usepackage{bbold}
\usepackage[footnotesize]{caption2}
\usepackage{graphicx}
\usepackage[center,footnotesize,hang]{subfigure}
\usepackage{url}
\usepackage{color}

\newcommand{\al}{\alpha}
\newcommand{\be}{\beta}

\newcommand{\De}{\Delta}

\newcommand{\vep}{\varepsilon}

\newcommand{\La}{\Lambda}
\newcommand{\om}{\omega}

\newcommand{\xit}{\tilde{\xi}}

 \def\cO{{\mathcal O}}

\newcommand{\phit}{\varphi_T}
\newcommand{\phis}{\varphi_S}

\newcommand{\diag}{\text{diag}}

\newcommand{\mean}[1]{\langle#1\rangle}

\newcommand{\ov}[1]{\overline{#1}}

\newcommand{\beq}{\begin{equation}}
\newcommand{\eeq}{\end{equation}}
\newcommand{\bac}{\beq\begin{array}}
\newcommand{\eac}{\end{array}\eeq}
\newcommand{\ba}{\begin{array}}
\newcommand{\ea}{\end{array}}
\newcommand{\bea}{\begin{eqnarray}}
\newcommand{\eea}{\end{eqnarray}}

\newcommand{\appendixA}{\setcounter{equation}{0}
\def\theequation{\rm{A}.\arabic{equation}}\section*}
\newcommand{\appendixB}{\setcounter{equation}{0}
\def\theequation{\rm{B}.\arabic{equation}}\section*}
\newcommand{\appendixC}{\setcounter{equation}{0}
\def\theequation{\rm{C}.\arabic{equation}}\section*}

\begin{document}
\begin{titlepage}
\vspace*{-1cm}
\phantom{hep-ph/***}

\hfill{DO-TH 10/21}

\hfill{TUM-HEP-776/10}

\vskip 1.5cm
\begin{center}
{\Large\bf Ultraviolet Completion of Flavour Models}
\vskip .3cm
\end{center}
\vskip 0.5  cm
\begin{center}
{\large Ivo de Medeiros Varzielas}\footnote{e-mail address: ivo.de@udo.edu}\\
\vskip .1cm
Fakult\"at f\"ur Physik, Technische Universit\"at Dortmund\\
D-44221 Dortmund, Germany\\
\vskip 1cm
{\large Luca Merlo}\footnote{e-mail address: merlo@tum.de}\\
\vskip .1cm
Physik-Department, Technische Universit\"at M\"unchen\\
James-Franck-Str. 1, D-85748 Garching, Germany\\
and\\
TUM Institute of Advanced Study\\ 
Lichtenbergstr. 2a, D-85748 Garching, Germany
\end{center}
\vskip 3cm
\begin{abstract}
\noindent
Effective Flavour Models do not address questions related to the nature of the fundamental renormalisable theory at high energies. We study the ultraviolet completion of Flavour Models, which in general has the advantage of improving the predictivity of the effective models. In order to illustrate the important features we provide minimal completions for two known $A_4$ models. We discuss the phenomenological implications of the explicit completions, such as lepton flavour violating contributions that arise through the exchange of messenger fields.
\end{abstract}
\end{titlepage}
\setcounter{footnote}{0}
\vskip2truecm

%
%

\topmargin -2.5 cm

\section{Introduction}

Model-building at the effective level is a very useful approach for beyond Standard Model (SM) physics. The non-renormalisable terms allowed by the symmetries are included in the Lagrangian, usually either up to the leading order (LO) or to the next to leading order (NLO). However, the effective models do not reveal what is the fundamental theory at high energy.  

The ultraviolet (UV) completion of effective models requires several additional messenger fields in order to produce the required effective Lagrangian through the combination of renormalisable terms. The messengers often play a role in the effective phenomenology of the model even if the explicit content is not specified (see for e.g. \cite{MR:SU3TB} where the messengers are responsible for the different hierarchies of the fermions despite unification). In turn, the explicit construction of a particular fundamental theory underlying the effective theory allows even greater control over the non-renormalisable terms - UV completions may not give rise to all the NLO terms of the effective theories particularly if they are minimal, where here we consider minimal those completions that have the least number of extra messenger fields and the least number of associated (renormalisable) couplings. As a result, we are motivated to build the underlying theory not simply due to philosophical considerations but because such a completion leads to improved predictivity (see e.g. \cite{KM:UVSO105d}).

For explicitness we consider two known models based on the $A_4$ discrete group, that deal only with the lepton sector and predict at LO the well-known Tri-Bimaximal (TB) mixing \cite{HPS:TBM,Xing:TBM}. The TB scheme is in good agreement with the neutrino oscillation global fits \cite{FLMPR:NuData1,STV:NuData,MS:NuData, FLMPR:NuData2,GMS:NuData,KamLAND:NuData}: the solar and the reactor angles are close to the $1\sigma$ value, while the atmospheric one is well inside this range. The deviations from TB mixing arise through corrections received by the vacuum and by the fermion mass matrices due to higher order terms in the respective Lagrangians.

We show that in minimal UV completions the NLO corrections to the vacuum can be entirely absent, with the LO vacuum preserved. The resulting lepton mixing is then exactly the TB pattern unless there is contamination between the neutrino and charged lepton sector. The two UV complete models we discuss here provide an example of each case: in the first model there is no contamination between the sectors, whereas in the second one this contamination gives rise to corrections to the mixing predicted at LO.

The introduction of new messengers can also mediate new lepton flavour violating (LFV) processes. We discuss this possibility in each of the UV complete models considered.

%
%
\section{The Altarelli-Feruglio Model}

The first model we consider is the original supersymmetric (SUSY) implementation of the Altarelli-Feruglio (AF) model \cite{AF:Extra,AF:Modular}. The flavour symmetry is a product of different terms:
\beq
G_f=A_4\times Z_3\times U(1)_{FN}
\eeq
where the spontaneous breaking of $A_4$ is responsible for the TB mixing, the cyclic symmetry $Z_3$ forbids dangerous couplings and helps to separate the charged lepton sector from the neutrino one, and the $U(1)_{FN}$ \cite{FN} provides a natural hierarchy among the charged lepton masses. \footnote{Instead of $U(1)_{FN}$ one can consider a $Z_N$ group with sufficiently high $N$. This avoids problems with gauged continuous groups with just one SM singlet scalar charged under the $U(1)_{FN}$ \cite{RV:Supergravity,MR:FSSUSYFP}.} $A_4$ is the group of the even permutations of $4$ objects (isomorphic to the group of discrete rotations in the three-dimensional space that leave invariant a regular tetrahedron) and it is a discrete subgroup of $SO(3)$. In Appendix A we present a detailed description of the $A_4$ group, including the multiplication rules of the irreducible representations.
The SUSY context simplifies the construction of the scalar potential, which provides a natural explanation for the vacuum. On the other hand, a continuous $R$-symmetry $U(1)_R$, that contains the usual $R$-parity as a subgroup, is added to $G_f$: under this symmetry the matter superfields transform as $U(1)_R=1$, while the scalar ones are neutral. As in the original paper, we do not discuss here the phenomenological consequences of the SUSY embedding.

Before we provide a minimal UV completion, we review the main features of the original model.

\subsection{The Effective Model}

Beyond the MSSM field content, the model contains new chiral superfields which are neutral under $SU(3)_c\times SU(2)_L \times U(1)_Y$, but transform under the flavour symmetry $G_f$: some of these new fields are neutral under $U(1)_R$ and are named flavons, while others transform as $U(1)_R=2$ and are named driving fields. With these charge assignments only the flavons can couple to the ordinary matter. 

In table \ref{table:AFtransformations} we show all the fields and their transformation properties under $G_f$.

\begin{table}[ht]
\begin{center}
\begin{tabular}{|c||ccccc||c||ccccc||ccc|}
\hline
&&&&&&&&&&&&&&\\[-4mm]
 & $\nu^c$ & $\ell$ & $e^c$ & $\mu^c$ & $\tau^c$ &$h_{u,d}$ & $\theta$ & $\phit$ & $\phis$ & $\xi$ & $\xit$ & $\phit^0$ & $\phis^0$ & $\xi^0$ \\[2mm]
\hline
&&&&&&&&&&&&&&\\[-4mm]
$A_4$ & $\bf3$ & $\bf3$ & $\bf1$ & $\bf1''$ & $\bf1'$ & $\bf1$ & $\bf1$ & $\bf3$ & $\bf3$ & $\bf1$ & $\bf1$ & $\bf3$ & $\bf3$ & $\bf1$ \\[2mm]
$Z_3$ & $\om^2$ & $\om$ & $\om^2$ & $\om^2$ & $\om^2$ & $1$ & $1$ & $1$ & $\om^2$ & $\om^2$ & $\om^2$ & $1$ & $\om^2$ & $\om^2$ \\[2mm]
$U(1)_{FN}$ & 0 & 0 & 2 & 1 & 0 & 0 & -1 & 0 & 0 & 0  & 0 & 0 & 0 & 0 \\[2mm]
$U(1)_R$ & 1 & 1 & 1 & 1 & 1 & 0 & 0 & 0 & 0 & 0  & 0 & 2 & 2 & 2 \\[2mm]
\hline
\end{tabular}
\end{center}
\caption{\it The transformation properties of the fields under $A_4$, $Z_3$, $U(1)_{FN}$ and $U(1)_R$.}
\label{table:AFtransformations}
\end{table}

The full superpotential of the theory at the LO accounts for three distinct types of contributions:
\beq
\label{eq:www}
w=w_ \ell +w_\nu+w_d
\eeq
where
\bea
\label{eq:AF_cl}
&&w_ \ell = \dfrac{y_e}{\Lambda^3}\theta^2e^c (\phit \ell) h_d + \dfrac{y_\mu}{\Lambda^2} \theta\mu^c(\phit \ell)^\prime h_d + \dfrac{y_\tau}{\Lambda}\tau^c(\phit \ell)^{\prime \prime} h_d\\
\label{eq:AF_nu}
&&w_\nu =y (\nu^c \ell)h_u+(x_A \xi+\tilde{x}_A \xit)(\nu^c\nu^c)+x_B(\phis\nu^c\nu^c)\phantom{\dfrac{\theta^2}{\Lambda^3}} \\
\label{eq:AF_d}
&&\begin{split}
w_d =&M (\phit^0 \phit) +g (\phit^0 \phit \phit) +\phantom{\dfrac{\theta^2}{\Lambda^3}} \\
& +g_1(\phis^0 \phis \phis) + g_2 \xit(\phis^0 \phis) +g_3 \xi^0 (\phis \phis) +g_4 \xi^0 \xi \xi + g_5\xi^0 \xi \tilde{\xi} +g_6 \xi^0 \tilde{\xi} \tilde{\xi} \;,
\end{split}
\eea
where we denote $(\ldots)\sim\bf1$, $(\ldots)'\sim\bf1'$ and $(\ldots)''\sim\bf1''$.
From $w_\ell$ and $w_\nu$ we can read the Yukawa interactions for charged leptons and neutrinos respectively, while from $w_d$ we obtain the vacuum expectation values (VEVs) for the scalar components of the flavon superfields (in the following we address as flavons only the scalar components of their respective superfields). Note that the driving superfields appear only linearly and as a result their scalar components do not acquire any VEV (this result is strictly true only in the SUSY exact phase \cite{FHM:Vacuum}). The stable vacuum which minimizes the scalar potential is given by
\beq
\dfrac{\langle\phit\rangle}{\Lambda}= (u,\,0,\,0)\;,\quad
\dfrac{\langle\phis\rangle}{\Lambda}=c_b(u,\,u,\,u)\;,\quad
\dfrac{\mean{\xi}}{\Lambda}=c_a\,u\,,\quad
\dfrac{\mean{\xit}}{\Lambda}=0\,,\quad
\dfrac{\mean{\theta}}{\La}=t\;,
\label{eq:AF_VEVs}
\eeq
where 
\beq
u=-\dfrac{3}{2}\dfrac{M}{g}\;,\qquad \qquad c_b^2=-\dfrac{g_4}{3g_3}c_a^2\;,\qquad\qquad c_a\text{ undetermined.}
\eeq
Some comments are in order. The scale $\La$ represents the cut-off of the theory and it will be substituted by the mass of the appropriate messenger fields in the UV completion of the model. Furthermore, the two triplets $\phit$ and $\phis$ develop VEVs in different directions in the flavour space: this is a key point of the model, because these directions define two distinct subgroups of the original $A_4$ group and it is this misalignment which gives rise to the TB mixing. The subgroup generated by $\phit$ is a $Z_3$, while that generated by $\phis$ is a $Z_2$ and these correspond also to the low-energy flavour structures of the charged lepton and neutrino mass matrices respectively. When the flavons acquire the VEVs in eq. (\ref{eq:AF_VEVs}) and the electroweak symmetry is broken by the Higgs VEVs $\mean{h_{u,d}}=v_{u,d}/\sqrt2$, the charged leptons develop a diagonal mass matrix and the neutrinos a mass matrix which can be diagonalized by the TB mixing:
\bea
&&m_\ell=\diag\left(y_e t^2,\,y_\mu t,\,y_\tau\right)\dfrac{v_d\, u}{\sqrt2}\;,
\label{eq:AF_ChargedMass}\\[2mm]
&&m_\nu=U_{TB}\diag\left(m_1,\,m_2,\,m_3\right)U_{TB}^T
\label{eq:AF_NuMass}
\eea
where the TB matrix is defined as
\beq
U_{TB}=\left(
         \begin{array}{ccc}
           \sqrt{2/3} & 1/\sqrt3 & 0 \\
           -1/\sqrt6 & 1/\sqrt3 & -1/\sqrt2 \\
           -1/\sqrt6 & 1/\sqrt3 & +1/\sqrt2 \\
         \end{array}
       \right)\;.
\label{eq:AF_TB}
\eeq
The corresponding lepton mixing angles are exactly those predicted by the TB pattern:
\beq
\sin^2\theta_{12}=\dfrac{1}{3}\;,\qquad\qquad \sin^2\theta_{23}=\dfrac{1}{2}\;,\qquad\qquad \sin\theta_{13}=0\;.
\eeq
As reported in the original papers the parameters $u$ and $t$ are restricted in well-defined ranges. Looking at the experimental values of the ratios among the charged lepton masses $t$ can be fixed around $0.05$. To constrain $u$ we should consider the expression of the $\tau$ mass, from which we get \cite{FHLM:LFVinSUSYA4}
\beq
u=\dfrac{\tan\beta}{|y_\tau|}\dfrac{\sqrt2 m_\tau}{v}\approx0.01\dfrac{\tan\beta}{|y_\tau|}\;,
\eeq
where $v\approx246$ GeV and $\tan\beta$ is the ratio between the VEVs of the neutral MSSM Higgses. If we require $|y_\tau|<3$, we find a lower bound on u of $0.05(0.007)$ for $\tan\beta=15(2)$. An upper bound on $u$ can be fixed by considering the NLO terms in the superpotential: the VEVs of the flavons and the mass matrices receive democratic corrections of order $u$ and as a result also the mixing angles deviate from the TB values of similar amounts; the largest value which does not spoil the agreement with the experimental data is taken from the allowed range of the solar angle and thus we see that $u$ can be at most $\sim0.05$. Therefore the ranges which $u$ and $t$ can span are given by
\beq
t\approx 0.05\;,\qquad\qquad 
0.001\lesssim u \lesssim 0.05\;.
\eeq

Apart from the masses and the mixing angles, the model predicts a well-defined relation among the $0\nu2\be$-decay effective mass and the lightest neutrino mass: here we report these predictions for the normal hierarchy (NH) and inverse hierarchy (IH) cases in terms of the neutrino mass squared differences $\De m^2_{atm}$ and $\De m^2_{sol}$
\beq
\ba{rl}
\mathrm{NH:}&|m_{ee}|=\dfrac{1}{3}\sqrt{\dfrac{9 m_1^4+2\De m^2_{atm}\De m^2_{sol}+m_1^2(10\De m^2_{atm}+\De m^2_{sol})}{m_1^2+\De m^2_{atm}}}\\[5mm]
\mathrm{IH:}&|m_{ee}|=\dfrac{1}{3m_3}\sqrt{9 m_3^4+m_3^2(8\De m^2_{atm}-\De m^2_{sol})+\De m^2_{atm}(-\De m^2_{atm}+\De m^2_{sol})}\;.
\ea
\eeq
These relations are valid only at the LO, as they receive corrections of order $u$ when the NLO terms are considered in the superpotential. 

The NLO contributions are on one hand welcome to explain the small deviation of the solar angle experimental central value from its TB value, and also to justify a possible non-vanishing value of the reactor angle as suggested in recent global fits on neutrino oscillations \cite{FLMPR:NuData1,STV:NuData,MS:NuData, FLMPR:NuData2,GMS:NuData,KamLAND:NuData}. On the other hand, they decrease the predictivity of the model by introducing uncertainties of order $u$ and then the model is more difficult to test. We will see in the next section that this last aspect is improved by considering the UV completion of the theory.

\subsection{The UV Completion}

The superpotential describing the neutrino sector is already fully renormalisable, with the effective neutrino masses resulting from the type I See-Saw mechanism. The driving superpotential that produces the required vacuum alignment is also renormalisable. 

The only effective terms we need to reproduce are those in the non-renormalisable superpotential $w_\ell$, which originates the charged lepton masses. In order to reproduce eq. (\ref{eq:AF_cl}), we minimally increase the field content introducing messengers $\chi_A$ and $\chi_A^c$, with $A=\tau,\,1,\,2,\,3$. In table \ref{table:AFtransformationsMessengers} we collect the transformation properties of all the messenger fields under the flavour symmetry $G_f$. Notice that these messengers must be chiral superfields with non-vanishing hypercharge: $-1$($+1$) for $\chi_A$($\chi_A^c$).

\begin{table}[ht]
\begin{center}
\begin{tabular}{|c||cccc||cccc|}
\hline
&&&&&&&&\\[-4mm]
 & $\chi_\tau$ & $\chi_1$ & $\chi_2$ & $\chi_3$ & $\chi^c_\tau$ & $\chi^c_1$ & $\chi^c_2$ & $\chi^c_3$ \\[2mm]
\hline
&&&&&&&&\\[-4mm]
$A_4$ & $\bf3$ & $\bf1'$ & $\bf1$ & $\bf1$ & $\bf3$ & $\bf1''$ & $\bf1$ & $\bf1$ \\[2mm]
$Z_3$ & $\om$ & $\om$ & $\om$ & $\om$ & $\om^2$ & $\om^2$ & $\om^2$ & $\om^2$ \\[2mm]
$U(1)_{FN}$ & 0 & 0 & 0 & -1 & 0 & 0 & 0 & +1 \\[2mm]
$U(1)_R$ & 1 & 1 & 1 & 1 & 1 & 1 & 1 & 1 \\[2mm]
\hline
\end{tabular}
\end{center}
\caption{\it The transformation properties of the messenger fields under $A_4$, $Z_3$, $U(1)_{FN}$ and $U(1)_R$.}
\label{table:AFtransformationsMessengers}
\end{table}

We can now write down the renormalisable charged lepton superpotential:
\beq
\label{eq:AF_ren}
w_\ell = M_{\chi_A} (\chi_A \chi^c_A) + h_d(\ell \chi^c_\tau) + \tau^c (\phit \chi_\tau)'' + \theta \mu^c\chi_1+ \theta e^c \chi_3 + (\phit \chi_\tau)' \chi^c_1 + (\phit\chi_\tau) \chi^c_2 + \theta\chi_2 \chi^c_3 \;.
\eeq
In order to keep the notation simple, we do not explicitly show the $\cO(1)$ coupling constants of each term. With this superpotential it is possible to construct the Feynman diagrams shown in appendix B from which the effective terms originate (after integrating out the messengers).
 
The introduction of the messengers could have some drawbacks. By introducing new fields in a theory, we concerns about possible mixings: in our minimal UV completion, however, the messengers do not share the same transformation properties with any other superfield and thus there are no mixings. Furthermore, looking at the eq. (\ref{eq:AF_ren}), we realize that the Lepton number is explicitly violated: in order to restore this global symmetry we could assign Lepton charge $-1$ ($+1$) to $\chi_A^c$ ($\chi_A$). However, since the Lepton number is already explicitly violated in the neutrino superpotential, we will not introduce this modification \footnote{If the heavy messengers have lepton number, they may be relevant to leptogenesis processes in the early Universe. Here we do not consider this possibility, even though it may provide further exceptions to the results of \cite{BBFN:FSLepto,ABMMM:TBLepto}.}.

Given the symmetry transformations in table \ref{table:AFtransformationsMessengers} we look for possible couplings which have not been required to produce the effective superpotential, but are allowed by the symmetry assignments nonetheless. These unwanted couplings could even rule out the model. In the proposed UV completion, there are three such couplings which we list in the following superpotential:
\beq
w_{new} =  \left(\phit \chi_\tau \chi^c_\tau\right) + \left(\phit\chi_\tau^c\right)''\chi_1 + \left(\phit\chi_\tau^c\right)\chi_2 \;.
\eeq
These three terms turn out to be innocuous. The first one simply redefines the mass terms of the messengers by relative factors of order $\cO(u)$ (in appendix B we report the corresponding Feynman diagram): as a consequence the charged lepton masses will be inversely proportional to $M_{\chi_\tau}(1+\cO(u))$. The other two terms in $w_{new}$ can only enable higher order diagrams much longer that those shown in appendix B, whose contributions can be absorbed in a redefinition of the renormalisable parameters.

Considering the total superpotential given by
\beq
w=w_ \ell + w_{new} + w_\nu + w_d
\eeq
we can express the effective coupling constants of eq. (\ref{eq:AF_cl}) in terms of those of the renormalisable ones. Looking at the ratios among the charged lepton masses and asking for $|y_\tau|<3$ we find the same constraint on $t$ and the same lower bound on $u$ as in the effective model.

It is interesting to investigate how the NLO terms described in \cite{AF:Modular} would be reproduced. Due to the symmetry content, the messengers we introduce in the minimal completion can not affect the driving or the neutrino superpotential. As a result the vacuum alignment of the flavon VEVs is preserved and the lepton mass matrices do not receive any correction: this can be understood considering that the $\phit$ sector is neutral under $Z_3$ while the $\phis$ one consists of fields which transform under $Z_3$, therefore there are no messengers which couple both sectors and it is not possible to generate corrections to the VEVs and to the existing flavour textures for charged leptons and neutrinos.

This fact has a deep impact on the model at the effective level and therefore it is interesting to consider it in more detail. One possibility is for example that an accidental symmetry is acting on the renormalisable superpotential. By extending the field content of the UV completion, we can introduce terms in the renormalisable driving superpotential that give rise to some NLO terms described in \cite{AF:Modular} at the effective level, and these should break such an accidental symmetry. These new messengers must have $U(1)_R=0,\,2$ and introduce mixing terms among $\phit$ and $\phis$. By inspection of all the new terms we can not identify any distinctive feature that reveals the accidental symmetry, and therefore we conclude that it is the minimal choice of the field content which prevents the appearance of any effective NLO term.

The absence of whole classes of NLO terms in the minimal completion is very relevant: without any higher order terms in the driving superpotential, the subgroups generated by $\phit$ and $\phis$ are preserved at NLO and beyond and there is no contamination between the sectors of the charged leptons ($\phit$) and of the neutrinos ($\phis$). The only new terms arise in the charged lepton superpotential, due to the presence of $w_{new}$ as already discussed above and they do not introduce any new flavour structure. As a result the charged lepton mass matrix keeps its diagonal form and we can conclude that the absence of any interaction among $\phit$ and $\phis$ prevents the contamination between the two sectors. There are no corrections to the lepton mixing angles, which are predicted to be the exact TB values. With respect to the effective theory, the UV completion can be easily tested looking at experimental deviations from the TB values of the mixing angles: if a discrepancy of more than $3\sigma$ will be pointed out, the minimal model should be ruled out and new ingredients, as new messengers, should be introduced to explain such observations. 

Due to the absence of any corrections to the mixing matrix, there is no CP violation in the lepton sector. As a result, leptogenesis cannot explain the present baryon asymmetry in the universe: in this case, we must look for either alternative baryogenesis mechanism or non-minimal completions of the effective AF model in order to switch on a CP violating phase. Finally, without any  corrections to the mixing matrix, the previously obtained upper bound on $u$ is lifted and it must now be constrained  by requiring that the charged lepton masses are stable: the flavon VEVs must be naturally smaller than the respective messenger masses to guarantee the correct mass hierarchy.

In family symmetry models, non-canonical K\"ahler terms are typically present but suppressed \cite{KPRV:Kahler}. In this UV completed model there are higher order effective terms in the K\"ahler potential, but they only give corrections to the existing canonical structures. Consider for example the construction of term that contribute to the $\ov{\ell} \ell$ K\"ahler term: we can construct an effective $\ov{\ell} \ell (\ov{\varphi}_T \phit)/ M^2$, which not only corresponds to a loop diagram suppressed by the factor $(\mean{\phit}/M_{\chi_\tau})^2$, but furthermore the flavour structure is unchanged from the tree-level canonical $\ov{\ell} \ell$.

Finally, as we have the model completely specified we can look at any possible rare-processes such as those giving rise to LFV: in principle we expect these contributions to also be suppressed. In the minimal completion proposed here we can not actually construct any LFV processes through the exchange of the messengers, because in the model each messenger is uniquely associated to a specific lepton flavour: e.g. $\chi^c_{\tau_2}$  always carries the $\tau$ flavour and $\chi_{\tau_3}$ always carries the $\tau^c$ flavour. There is no coupling that can enable a conversion into different components of the lepton $SU(2)_L$ doublet $\ell$, even in higher order diagrams. As a result the analysis on the LFV processes should be carried out following \cite{FHLM:LFVinA4,FHLM:LFVinSUSYA4,HMP:LFVinSUSYA4SS,FHM:Vacuum,FP:RareDecaysA4} but considering the new flavour structures of the mass matrices.

%
%
\section{The Altarelli-Meloni Model}

The second model we consider is the Altarelli-Meloni (AM) model \cite{AM:A4xZ4}. It is very similar to the AF realization, but it combines the $U(1)_{FN}$ and $Z_3$ terms of the complete flavour symmetry in a single $Z_4$ group. This reduces the symmetry and field content of the model (for a similar approach see also \cite{Lin:Predictive}). The complete flavour symmetry is then given by: 
\beq
G_f=A_4\times Z_4\;.
\eeq
The $A_4$ terms behave very similarly to what was already described in the AF model, although note that the VEV of the $\phit$ field is now in the  $(0,1,0)$ direction. The $Z_4$ substitutes the $Z_3$ in preventing dangerous couplings and also substitutes the $U(1)_{FN}$ in providing the correct charged lepton mass hierarchy. Once again a continuous $R$-symmetry $U(1)_R$ is used for the SUSY implementation.

We again precede our proposal for a minimal UV completion with a brief review of the effective model.

\subsection{The Effective Model}

As in the AF realization, apart from the usual MSSM superfields the spectrum includes flavons and driving fields. In table \ref{table:AMtransformations} we show the fields and their transformation properties under the symmetries.

\begin{table} [h]
\centering
\begin{tabular}{|c||c|c|c|c|c||c|c||c|c|c|c||c|c|c|c|}
\hline
&&&&&&&&&&&&&&\\[-4mm]
{\tt Field}& $\nu^c$ & $\ell$ & $e^c$ & $\mu^c$ & $\tau^c$ & $h_d$ & $h_u$& 
$\varphi_T$ &  $\xi'$ & $\varphi_S$ & $\xi$ & $\varphi_0^T$  & $\varphi_0^S$ & $\xi_0$\\[2mm]
\hline
&&&&&&&&&&&&&&\\[-4mm]
$A_4$ & $3$ & $3$ & $1$ & $1$ & $1$ & $1$ &$1$ &$3$ & $1'$ & $3$ & $1$ &  $3$ &  $3$ & $1$\\[2mm]
$Z_4$ & -1 & i & $1$ & i & -1 & 1 & i & i & i  & $1$ & $1$ & -1 &  $1$ & $1$\\[2mm]
$U(1)_R$ & $1$& $1$ & $1$ & $1$ & $1$ & $0$ & $0$ & $0$& $0$  & $0$ & $0$ & $2$ & $2$ & $2$\\[2mm]
\hline
\end{tabular}
\caption{\it Transformation properties of all the fields of the model under $A_4$, $Z_4$ and $U(1)_R$.}
\label{table:AMtransformations}
\end{table}
The superpotential is divided according to sectors as in eq. (\ref{eq:www}).
The neutrino and the driving superpotentials are renormalisable,
\bea
&&w_\nu = y_\nu (\nu^c \ell)\,h_u + (M+a\,\xi)\,\nu^c\nu^c+b \,\nu^c\nu^c\,\varphi_S
\label{AM_nu}\\[2mm]
&&\begin{split}
w_d=&M (\varphi_0^S \varphi_S)+g_1 (\varphi_0^S \varphi_S\varphi_S)+
g_2 \xi (\varphi_0^S \varphi_S)+
g_3 \xi_0 (\varphi_S\varphi_S)+
g_4 \xi_0 \xi^2+
M_\xi \xi_0 \xi + \\
 &+  M_0^2 \, \xi_0 +
h_1 \xi' (\varphi_0^T \varphi_T)''+
h_2 (\varphi_0^T \varphi_T\varphi_T)\;,
\end{split}
\label{AM_driving}
\eea
whereas the charged lepton superpotential has the effective terms
\beq
\begin{split}
{w_\ell}=&\hspace{5mm}\dfrac{y_\tau}{\Lambda} \tau^c (\varphi_T \ell ) \, h_d + \\
&+\dfrac{y_\mu}{\Lambda^2} \mu^c (\varphi_T \varphi_T\ell ) \, h_d +
\dfrac{y_\mu'}{\Lambda^2} \mu^c (\varphi_T\ell )^{''} \xi' \, h_d +\\
&+\dfrac{y_e}{\Lambda^3} e^c (\varphi_T\varphi_T\ell )^{''} \xi' \, h_d +
\dfrac{y_e'}{\Lambda^3} e^c (\varphi_T\ell )' \xi^{'2} \, h_d +
\dfrac{y_e''}{\Lambda^3} e^c (\varphi_T\ell )' (\varphi_T \varphi_T )'' \, h_d + \\
&+\dfrac{y_e'''}{\Lambda^3} e^c (\varphi_T\ell )'' (\varphi_T \varphi_T )' \, h_d +
\dfrac{y_e^{\rm iv}}{\Lambda^3} e^c (\varphi_T\ell ) (\varphi_T \varphi_T ) \, h_d\;.
\end{split}
\label{AM_cl}
\eeq
In the previous equation, all the couplings contributing to the same mass term are of the same order of magnitude (when not vanishing) once the flavour symmetry is spontaneously broken. As reported in \cite{AM:A4xZ4} the minimum of the scalar potential in the exact SUSY limit is given at the LO by
\beq
\dfrac{\langle\phit\rangle}{\Lambda}= (\vep,\,0,\,0)\;,\quad
\dfrac{\mean{\xi'}}{\Lambda}=c\,\vep\,,\quad
\dfrac{\langle\phis\rangle}{\Lambda}=(\vep',\,\vep',\,\vep')\;,\quad
\dfrac{\mean{\xi}}{\Lambda}=c'\,\vep'\,,
\label{eq:AM_VEVs}
\eeq
where 
\bea
&&c=-2\dfrac{h_2}{h_1}\;,\qquad \qquad \vep\text{ undetermined,}\\[2mm]
&&c'=-\dfrac{M}{g_2\vep'}\;,\qquad \quad\;\; \vep^{\prime 2}=\dfrac{1}{3g_2^2g_3}\left(g_2(M M_\xi-g_2 M_0^2)-g_4 M^2\right)\;.
\eea
Notice that the VEVs of $\phit$ and $\phis$ break the flavour symmetry into different directions: as in the AF model $\phis$ breaks $A_4$ down to its $Z_2$ subgroup; on the other hand, $\phit$ is aligned along a direction which breaks $A_4$ completely. It is interesting to note that this direction  preserves a $Z_3$ subgroup of $A_4\times Z_3$ as an accidental symmetry of the LO superpotential.

When the electroweak and the flavour symmetries are spontaneously broken, the charged lepton mass matrix becomes
\beq
m_\ell=\diag\left(\cO(\vep^2),\,\cO(\vep),\,\cO(1)\right)\dfrac{v_d\,\vep}{\sqrt2}\;,
\label{eq:AM_ChargedMass}
\eeq
while the neutrino mass matrix can be exactly diagonalized by the TB mixing. The experimental charged lepton mass hierarchy can be fitted if $\vep$ is close to the square of the Cabibbo angle,
\beq
\vep\approx \cO(\lambda_C^2)\;.
\eeq

When considering the NLO corrections as in \cite{AM:A4xZ4}, the LO predictions are modified by terms of order $\vep'$ and as a result
\beq
\sin^2\theta_{12}=\dfrac{1}{3}+\cO(\vep')\;,\qquad\qquad \sin^2\theta_{23}=\dfrac{1}{2}+\cO(\vep')\;,\qquad\qquad \sin\theta_{13}=\cO(\vep')\;.
\label{AM_NLOAngles}
\eeq
As in the AF model, we can place an upper bound on $\vep'$ by considering the solar angle: 
\beq
\vep'\lesssim\cO(\lambda_C^2)\;.
\eeq

\subsection{The UV Completion \label{AM_UV}}

We can get the effective terms from eq. (\ref{AM_cl}) at the renormalisable level by adding the messengers. In Table \ref{table:AMtransformationsMessengers} we report their charge assignments. As in the AF model, $\chi_A (\chi^c_A)$ ($A$ is $\tau$, $1, 2, 3$) have $-1(+1)$ hypercharge. 

\begin{table}[ht]
\begin{center}
\begin{tabular}{|c||cccc||cccc|}
\hline
&&&&&&&&\\[-4mm]
 & $\chi_\tau$ & $\chi_1$ & $\chi_2$ & $\chi_3$ & $\chi^c_\tau$ & $\chi^c_1$ & $\chi^c_2$ & $\chi^c_3$ \\[2mm]
\hline
&&&&&&&&\\[-4mm]
$A_4$ & $\bf3$ & $\bf1''$ & $\bf1'$ & $\bf1''$ & $\bf3$ & $\bf1'$ & $\bf1''$ & $\bf1'$ \\[2mm]
$Z_4$ & $i$ & $-1$ & $-1$ & $-i$ & $-i$ & $-1$ & $-1$ & $i$ \\[2mm]
$U(1)_R$ & 1 & 1 & 1 & 1 & 1 & 1 & 1 & 1 \\[2mm]
\hline
\end{tabular}
\end{center}
\caption{\it The transformation properties of the messenger fields under $A_4$, $Z_3$ and $U(1)_R$.}
\label{table:AMtransformationsMessengers}
\end{table}

With the chosen field content the renormalisable terms of the charged lepton sector are
\beq
\label{eq:AM_ren}
w_\ell =M_{\chi_A} (\chi^c_A \chi_A) + h_d(\ell \chi^c_\tau) + \tau^c (\phit \chi_\tau) + \mu^c \xi' \chi_1+ e^c \xi' \chi_3 +(\phit \chi_\tau)'' \chi^c_1 +(\phit \chi_\tau)'  \chi^c_2 +  \chi^c_3 \xi' \chi_2 \,.
\eeq 
Some comments are in order. Note that these renormalisable terms do not generate all of the non-renormalisable terms of eq. (\ref{AM_cl}): for example the only $e^c$ term we reproduce is the $y_e'$ term (with two insertions of $\xi'$). This however does not affect the low-energy observables, as all the entries are produced at the same order as in the original effective theory. Notice that due to the alignment of the VEV of $\phit$, the messengers carrying each flavour are changed with respect to the AF completion - for example the same $(\phit \chi_\tau)$ term assigns messenger $\chi_{\tau_3}$ to the $\tau^c$ flavour (instead of $\chi_{\tau_1}$ as in the AF model). Furthermore, just as in the AF completion there are a few extra terms:
\beq
w_{new} = (\phis\chi_\tau \chi^c_\tau) + \xi(\chi_A \chi^c_A) \,.
\eeq
As a result, each pair of messengers (including the triplets) will acquire a $\langle \xi \rangle$ correction to their mass term, which are however aligned with the previous contributions.
Distinctly unlike what happens in the AF completion, $w_{new}$ introduces new flavour structures into the resulting effective model through the accidental $(\phis\chi_\tau \chi^c_\tau)$ term - which can not be avoided even by redefining the $Z_4$ charges as $\phis$ must be a $Z_4$ singlet (this requirement on the charge of $\phis$ is imposed by the required driving terms, which include simultaneously linear terms in $\phis$ and quadratic terms in $\phis$).
This accidental term contaminates the charged lepton sector with the field that acquires the vacuum used in the neutrino sector, and therefore we generally expect corrections to the TB mixing. This occurs even though the vacuum is preserved at NLO and beyond (the symmetry content forbids the messengers from interfering with the driving sector, as in the AF completion). The contamination term also enables other processes like LFV decays, that could not be present in the AF completion. Finally, corrections to the existing mass structures arise from diagrams involving the respective vertex, presented in the appendix C.

We can account for the contamination by noting that the term modifies the standard $\chi_\tau$ mass matrix
\beq
M^{LO}_{\chi_\tau}=M_{\chi_{\tau}} \left(
         \begin{array}{ccc}
           1 & 0 & 0 \\
           0 & 0 & 1 \\
           0 & 1 & 0 \\
         \end{array}
       \right)
\label{eq:Mct_LO}
\eeq
by adding
\beq
M^{NLO}_{\chi_\tau}=\vep' M_{\chi_{\tau}}\left(
         \begin{array}{ccc}
           c_\xi+2 c_s & -c_s+c_a & -c_s-c_a \\
           -c_s-c_a & 2c_s & c_\xi-c_s+c_a \\
           -c_s+c_a & c_\xi-c_s-c_a & 2c_s \\
         \end{array}
       \right)\;,
\label{eq:Mct_NLO}
\eeq
where $c_\xi$, $c_s$ and $c_a$ are $\cO(1)$ coefficients which define the $\xi(\chi_\tau \chi^c_\tau)$ term, the symmetric and the antisymmetric $(\phis\chi_\tau \chi^c_\tau)$ products respectively. These NLO contributions enable off-diagonal corrections in the charged lepton mass matrix that are proportional to the LO masses of the charged leptons with a small parameter that depends on the appropriate mass insertion that enables it
\beq
m^{NLO}_{\ell}=\left(
         \begin{array}{ccc}
           m_e & \epsilon\, m_e & \epsilon'\, m_e \\
           \epsilon\, m_\mu & m_\mu & \epsilon''\, m_\mu \\
           \epsilon'\, m_\tau & \epsilon''\, m_\tau & m_\tau \\
         \end{array}
       \right)\;,
\label{eq:mcl_NLO}
\eeq
where $m_{e,\mu,\tau}$ can be read from eq. (\ref{eq:AM_ChargedMass}) and the $\epsilon$, $\epsilon'$, $\epsilon''$ coefficients  are $\cO(\vep')$ factors that can be written in terms of $c_\xi$, $c_s$, $c_a$ and $\vep'$. In particular, note that the $\epsilon$ entries are mediated by the $12$ or $31$ entries of $M^{NLO}_{\chi_\tau}$, the $\epsilon'$ by the $13$ or $21$ entries and the $\epsilon''$ entries by the $22$ and $33$. One can obtain the diagram that leads to $\epsilon\, m_e$ ($\epsilon\, m_\mu$) starting from the term that would generate the standard electron (muon) mass with a mass insertion contained in $\left(M^{NLO}_{\chi_\tau}\right)_{ij}$, which causes the diagram to finish with a $\mu^c$ ($e^c$). This way it is possible to understand why the same parameter is present in two off-diagonal elements of the charged lepton mass matrix (here the same $\epsilon$ as they are both mediated by the combination $-c_s+c_a$).
Note also that it is the $e^c$, $\mu^c$ or $\tau^c$ present in the diagram that controls the overall suppression, as it specifies the number of $\xi'$ insertions needed - which is why (in our notation) each row is respectively proportional to $m_e$, $m_\mu$ and $m_\tau$.

The contamination discussed above does not affect the neutrino sector and as a result the neutrino mass matrix is exactly diagonalized by the TB mixing. The corrections to the charged lepton mass matrix switch on deviations in the mixing matrix and the resulting mixing angles are those described by eq. (\ref{AM_NLOAngles}).

Once again there are higher order K\"ahler terms which are suppressed, but unlike what occurs in the AF completion there are non-canonical contributions enabled by the contamination term that present a different flavour structure with respect to the canonical terms. An example is an effective $\ov{\ell} \ell (\ov{\varphi}_S \phis)/ M^2$ term which corresponds to a loop diagram suppressed by order $(\mean{\phis}/M_{\chi_\tau}) ^2$ (the diagram is presented in the appendix C).

In this contaminated completion, a given messenger is no longer uniquely associated to a specific lepton flavour. The exclusive association of messengers to a given flavour is lost at NLO due to the $\phis$ term. One example of a LFV $\ell_j$ to $\ell_k$ conversion process mediated by the messengers is a loop diagram where the components of $\chi^c_\tau$ and $\chi_\tau$ are exchanged inside the loop by having a $\phis$ insertion. The loop can be closed by changing back through the LO mass term with the end result of starting on the left with a given generation (a given flavour of $\ell$ goes into a specific messenger $\chi^c_{\tau_j}$ according to the $A_4$ contraction) and finishing on the right with a different one ($\chi^c_{\tau_j}$ changes non-canonically into $\chi_{\tau_k}$ and eventually finishes with a different $\ell$ flavour) with the emission of a photon: the diagram is presented in the appendix C, where for simplicity we denoted the index of the lepton to be the same as the index of the messenger (the appropriate $A_4$ invariant contraction is understood). In order to estimate this process, it is convenient to use a mass insertion approach where we consider the $\phis$ effect as non-canonical entries in the mass matrix of $\chi^c_{\tau}$, $\chi_\tau$ (see eq.(\ref{eq:Mct_NLO})). As a final result we find that the normalized branching ratio of the described process is given by
\beq
BR(\ell_j\rightarrow \ell_k\gamma)\approx \dfrac{48\pi^3\alpha}{G_F^2}\left(2.5\times10^{-4}\right)\left(\dfrac{(M^{NLO}_{\chi_\tau})_{jk}}{M^3_{\chi_\tau}}\right)^2\;,
\eeq
where $G_F$ is the Fermi constant and $\alpha$ the fine structure constant. Considering the present upper bound on $BR(\mu\rightarrow e\gamma)=1.2\times10^{-11}$, it corresponds to an upper bound on the masses of the messengers of about 
\beq
M_{\chi_\tau}\approx 10^4  \text{  GeV}\;.
\eeq
This shows that these processes are strongly suppressed when considering realistic messenger masses.

%
%
\section{Conclusion}

Specifying the ultraviolet completion of flavour models can significantly increase the predictivity of the theory, eliminating several terms that would be included at the effective level. This is particularly true of minimal completions where attention is given to minimise the number of messengers fields and associated couplings. We exemplified this claim by providing minimal completions of two distinct $A_4$ lepton flavour models. In particular we find that although at the effective level the vacuum would be expected to receive significant next to leading order corrections, in the ultraviolet complete theory the vacuum obtained at leading order can turn out to be preserved. 

In the ultraviolet completion of the effective Altarelli-Feruglio model the vacuum of flavon fields is indeed preserved and no corrections to the lepton mass matrices are switched on. As a result, Tri-Bimaximal mixing is exactly predicted and in particular the reactor angle is exactly vanishing; no CP violating phase arises in this context and then leptogenesis cannot occur to explain the baryon asymmetry of the universe; non-canonical kinetic terms are absent and furthermore new lepton flavour violating transitions mediated by the messengers do not appear.

On the other hand, in the ultraviolet completion of the Altarelli-Meloni model there is a contamination effect between the charged lepton and the neutrino sectors. This is a very important distinction and a key aspect in this model, as it is through the contamination that interesting phenomenological consequences are enabled at the next to leading order (even if suppressed). The end results of the contamination include non-canonical K\"ahler terms, lepton flavour violating processes and also corrections to the leading order prediction of Tri-Bimaximal lepton mixing - which in particular can account for non-vanishing values of the reactor angle.

%
%

\section*{Acknowledgements}

We thank Guido Altarelli and Paride Paradisi for useful comments. IdMV acknowledges hospitality at TU M\"unchen during a part of this work. IdMV was supported by DFG grant PA 803/6-1.

%
%
\newpage
\mathversion{bold}
\appendixA{Appendix A: The Group $A_4$}
\mathversion{normal}

The group $A_4$ is generated by two elements $S$ and $T$ obeying the relations \cite{GR_Hamermesh,GR_Cornwell}:
\beq
S^2=(ST)^3=T^3=1\;.
\eeq
It has three independent one-dimensional representations, $\bf1$, $\bf1'$ and $\bf1''$ and one three-dimensional representation $\bf3$.
The one-dimensional representations are given by:
\beq
\begin{array}{lll}
{\bf1} & S=1 & T=1 \\[3mm]
{\bf1'} & S=1 & T=e^{i 4 \pi/3} \equiv \omega^2\\[3mm]
{\bf1''}& S=1 & T=e^{i 2\pi/3} \equiv\omega\\[3mm]
\end{array}
\eeq
The three-dimensional representation, in a basis where the generator $T$ is diagonal, is given by:
\beq
T=\left(
        \begin{array}{ccc}
        1 & 0 & 0 \\
        0 & \omega^2 & 0 \\
        0 & 0 & \omega \\
        \end{array}
    \right),\qquad\qquad
S=\dfrac{1}{3}
    \left(
        \begin{array}{ccc}
        -1 & 2 & 2 \\
        2 & -1 & 2 \\
        2 & 2 & -1 \\
        \end{array}
    \right)\;.
\label{ST}
\eeq

We now report the multiplication rules between the various representations. In the following we use $\alpha=(\alpha_1,\,\alpha_2,\,\alpha_3)$ to indicate the elements of the first representation of the product and $\beta=(\beta_1,\,\beta_2,\,\beta_3)$ to indicate those of the second representation. Moreover $a,b=0,\pm1$ and we denote ${\bf1}^0\equiv{\bf1}$, ${\bf1}^1\equiv{\bf1}^\prime$, ${\bf1}^{-1}\equiv{\bf1}^{\prime\prime}$ and similarly for the doublet representations. On the right-hand side the sum $a+b$ is modulo 3.

We start with all the multiplication rules which include the one-dimensional representations:
\beq
\begin{array}{l}
{\bf1}\times {\bf r}={\bf r}\times{\bf1}={\bf r}\qquad\text{with ${\bf r}$ any representation}\;,\\[3mm]
{\bf1}^a\times{\bf1}^b={\bf1}^b\times{\bf1}^a={\bf1}^{a+b}\sim\alpha\beta\;,\\[3mm]
{\bf1}^\prime\times{\bf3}={\bf3}\sim\left(\begin{array}{c}
                                        \alpha\beta_3 \\
                                        \alpha\beta_1 \\
                                        \alpha\beta_2\\
                                \end{array}\right)\;,\qquad
{\bf1}^{\prime\prime}\times{\bf3}={\bf3}\sim\left(\begin{array}{c}
                                        \alpha\beta_2 \\
                                        \alpha\beta_3 \\
                                        \alpha\beta_1\\
                                \end{array}\right)\;.
\end{array}
\eeq

The multiplication rule with the three-dimensional representation is
\beq
\begin{array}{l}
{\bf3}\times{\bf3}={\bf3}_S+{\bf3}_A+{\bf1}+{\bf1}'+{\bf1}''\quad\text{with}\quad\!\left\{
        \begin{array}{l}
        {\bf1}\;\,\sim\alpha_1\beta_1+\alpha_2\beta_3+\alpha_3\beta_2\;,\\[3mm]
        {\bf1}'\;\sim\alpha_3\beta_3+\alpha_1\beta_2+\alpha_2\beta_1\;,\\[3mm]
        {\bf1}''\,\sim\alpha_2\beta_2+\alpha_1\beta_3+\alpha_3\beta_1\;,\\[3mm]
        {\bf3}_S\sim\dfrac{1}{3}\left(\begin{array}{c}
                                     2\alpha_1\beta_1-\alpha_2\beta_3-\alpha_3\beta_2\\
                                     2\alpha_3\beta_3-\alpha_1\beta_2-\alpha_2\beta_1\\
                                     2\alpha_2\beta_2-\alpha_1\beta_3-\alpha_3\beta_1\\
                                     \end{array}
                               \right)\\[3mm]
        {\bf3}_A\sim\dfrac{1}{2}\left(\begin{array}{c}
                                 \alpha_2\beta_3-\alpha_3\beta_2\\
                                 \alpha_1\beta_2-\alpha_2\beta_1\\
                                 \alpha_3\beta_1-\alpha_1\beta_3\\
                                \end{array}\right)
\end{array}\right.
\ea
\eeq
Note that due to the choice of complex representation matrices for the real representation ${\bf3}$ the conjugate $\al^*$ of $\al \sim {\bf3}$ does not transform as ${\bf3}$, but rather $(\al_1^{\star},\, \al_3^*,\, \al_2^*)$ transforms as triplet under $A_4$. The reason for this is that $T^*= U_{23}^T\,T\,U_{23}$ and $S^*=U_{23}^T\,S\,U_{23}=S$ where $U_{23}$ is the matrix which exchanges the 2nd and 3rd row and column.

%
%
\appendixB{Appendix B: Diagrams for the AF Model}

Here we provide the diagrams which generate the effective terms in the AF model. In order to provide the correct diagrams, we consider first the diagram with the superfield and then we perform the projection in the SM field content leading to the diagrams shown here. The diagrams in fig. \ref{AF_Diagrams}(a,b,c) originate the tau, mu and electron mass terms of eq. (\ref{eq:AF_cl}). The diagram in fig. \ref{AF_Diagrams}(d) represents a generic modification of the previous graphs due to the presence of the term $\left(\phit \chi_\tau \chi^c_\tau\right)$ in the renormalisable superpotential. These last contributions term are however innocuous because they can be absorbed in the definition of the charged lepton masses. Notice that the repeated indices are symbolic of the $A_4$ invariant contraction.

\begin{figure}[h!]
 \centering
\subfigure[Mass term of the $\tau$ lepton.]
   {\includegraphics[height =3.5cm]{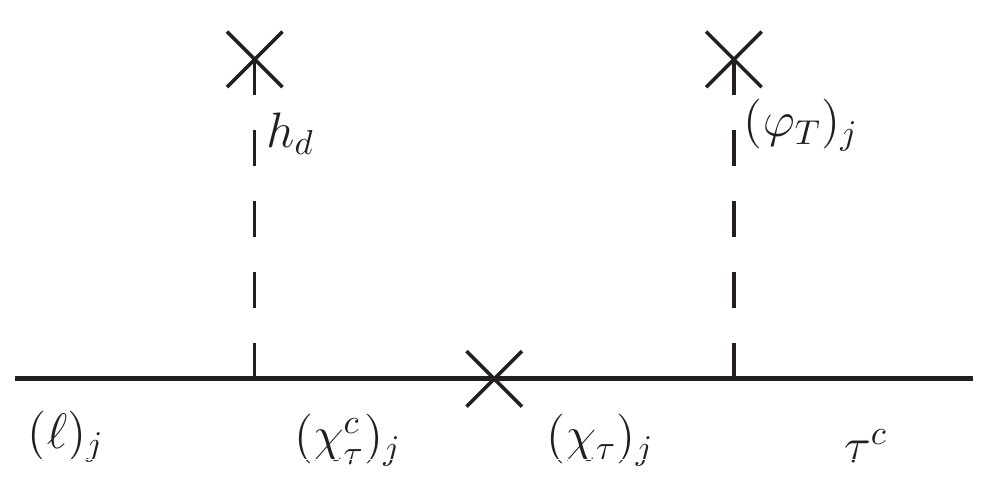}}
\subfigure[Mass term of the $\mu$ lepton.]
   {\includegraphics[height =3.5cm]{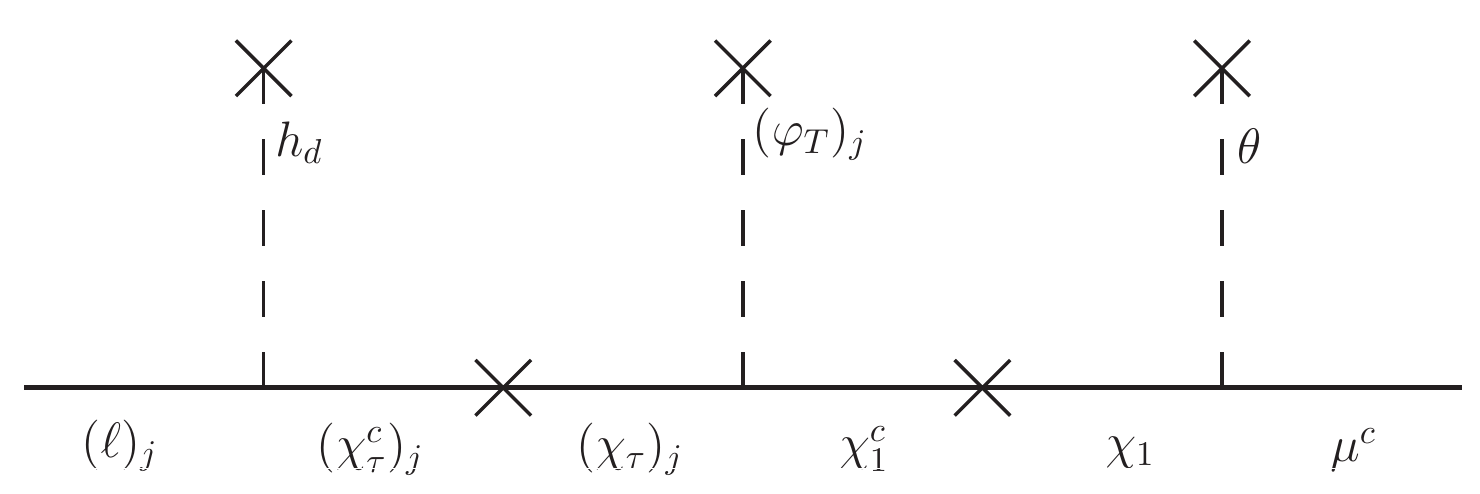}}
\subfigure[Mass term of the $e$ lepton.]
   {\includegraphics[height=3.5cm]{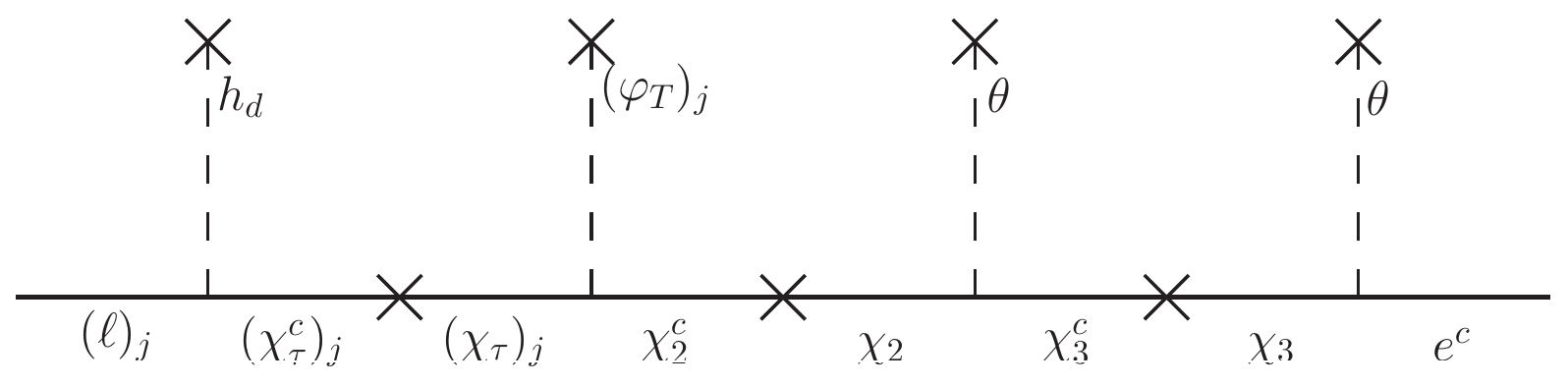}}
\subfigure[General corrections to the mass terms.]
   {\includegraphics[height=3.5cm]{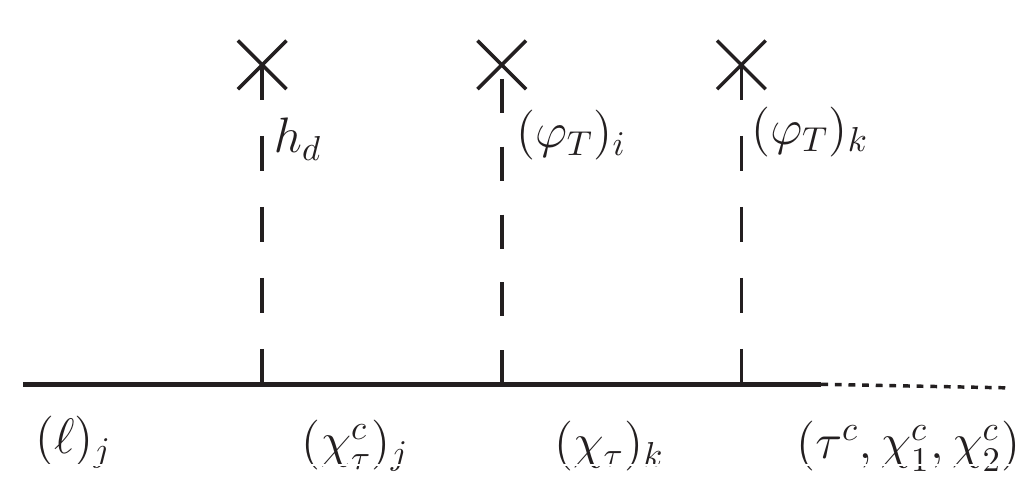}}
\caption{Diagrams for the effective mass terms and their corrections.}
\label{AF_Diagrams}
\end{figure}

%
%
\appendixC{Appendix C: Diagrams for the AM Model}

In this appendix we show the relevant diagrams for the AM model. The diagram which originates the tau mass term of eq. (\ref{AM_cl}) is the same of fig. \ref{AF_Diagrams}(a) and therefore we do not repeat it here. The diagrams in fig. \ref{AM_Diagrams}(a,b) originate the mu and electron mass terms of eq. (\ref{AM_cl}). The diagrams in fig. \ref{AM_Diagrams}(c,d) represent the modifications of the previous graphs due to the presence of $\left(\phis \chi_\tau \chi^c_\tau\right)$ and $\xi\left(\chi_\tau \chi^c_\tau\right)$ in the renormalisable superpotential. While the second of these terms generates contributions which can be absorbed in a redefinition of the diagonal LO charged lepton masses, the first term introduces a new flavour structure: it switches on all the off-diagonal entries of the charged lepton mass matrix. The phenomenological consequences have been discussed in section \ref{AM_UV}. Notice that the repeated indices implicilty represent the appropriate $A_4$ invariant contraction.

\begin{figure}[h!]
 \centering
\subfigure[Mass term of the $\mu$ lepton.]
   {\includegraphics[height =3.5cm]{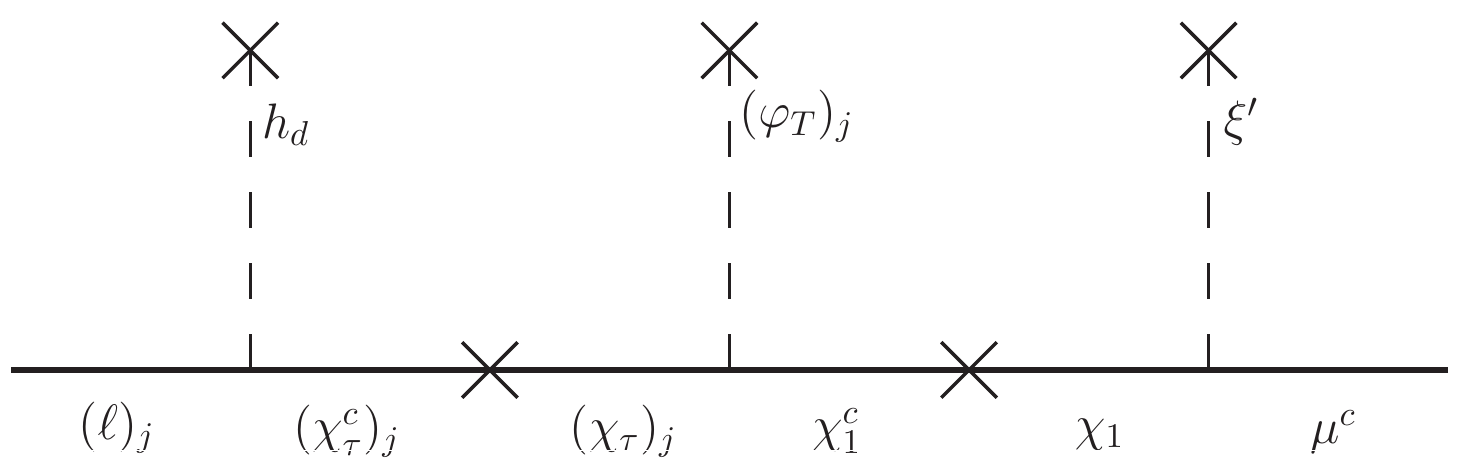}}
\subfigure[Mass term of the $e$ lepton.]
   {\includegraphics[height=3.5cm]{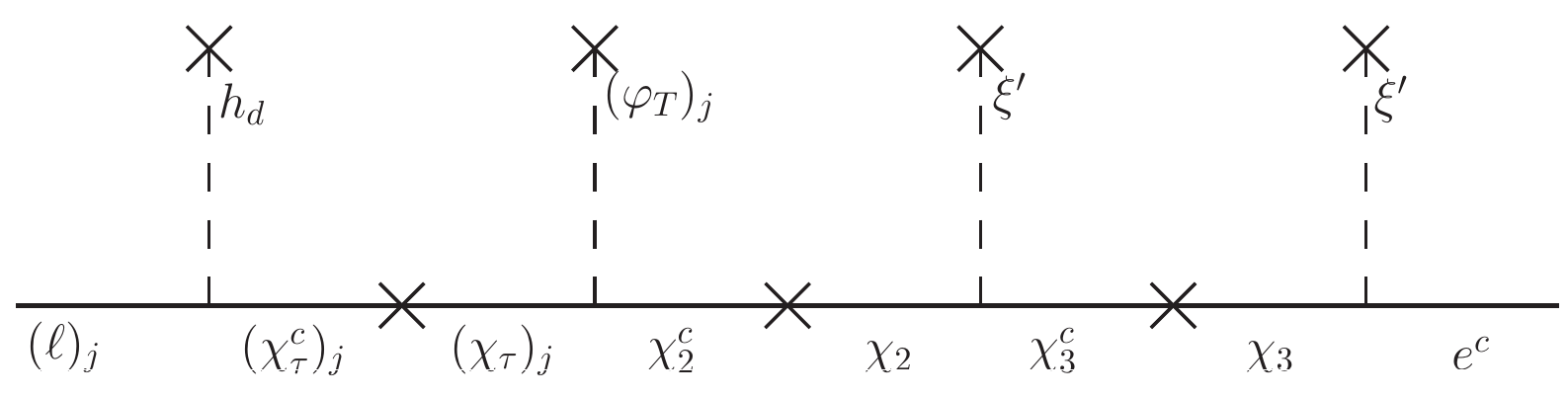}}
\subfigure[General corrections to the mass terms with the $\phis$ insertion.]
   {\includegraphics[height=3.5cm]{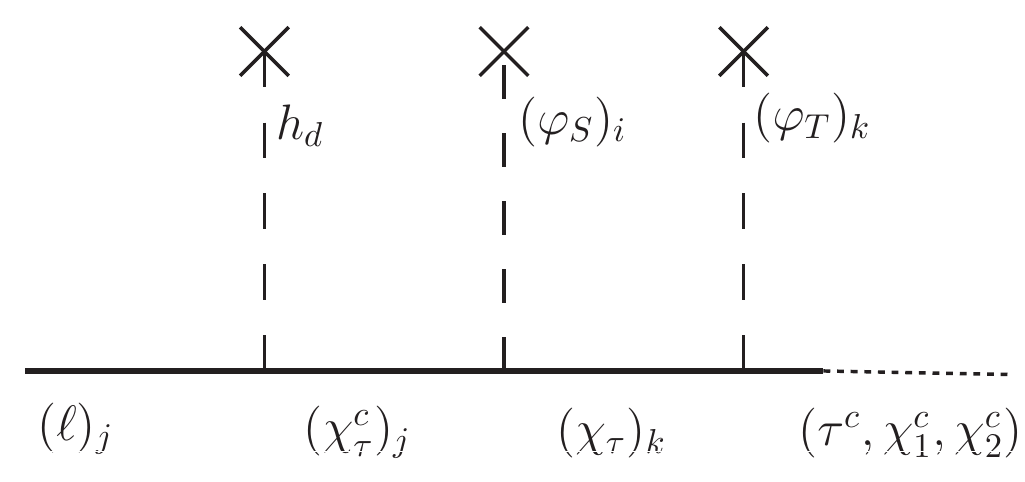}}
\subfigure[General corrections to the mass terms with $\xi$ insertion.]
   {\includegraphics[height=3.5cm]{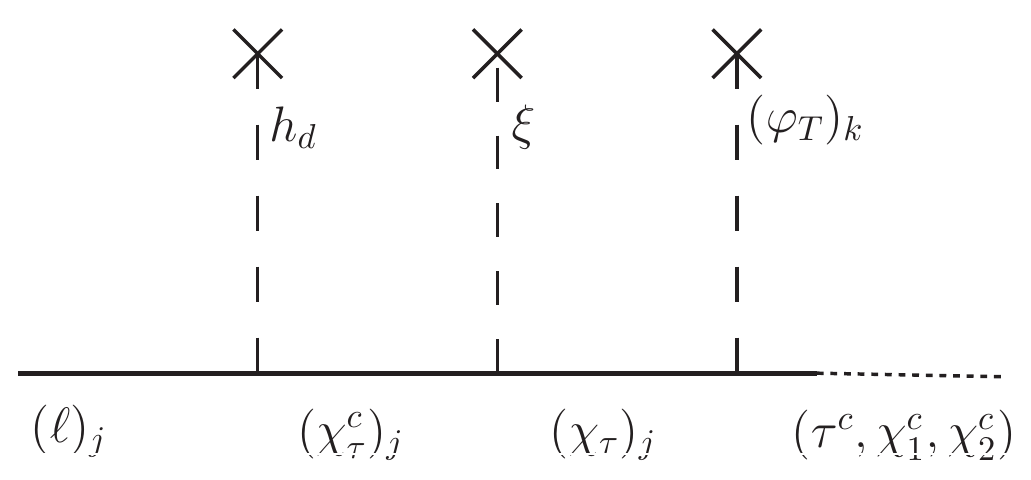}}
\caption{Diagrams for the effective mass terms and their corrections.}
\label{AM_Diagrams}
\end{figure}

Finally in fig. (\ref{AM_Kinetic}) we report the diagram which is responsible for the non-canonical effective K\"ahler term $\ov{\ell} \ell (\ov{\varphi}_S \phis)/ M^2$ and the diagram responsible for LFV process $\ell_j \rightarrow \ell_k \gamma$.

\begin{figure}[h!]
 \centering
 \subfigure[Non-canonical effective kinetic term.]
   {\includegraphics[height =6cm]{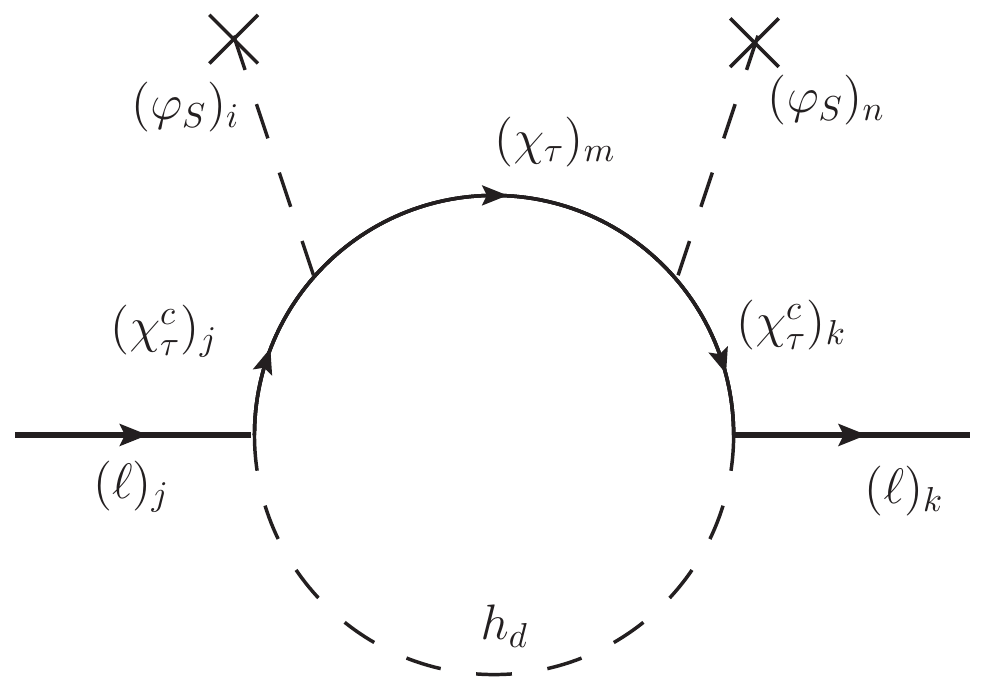}}
 \subfigure[ $\ell_j\rightarrow \ell_k\gamma$ process.]
   {\includegraphics[height =8cm]{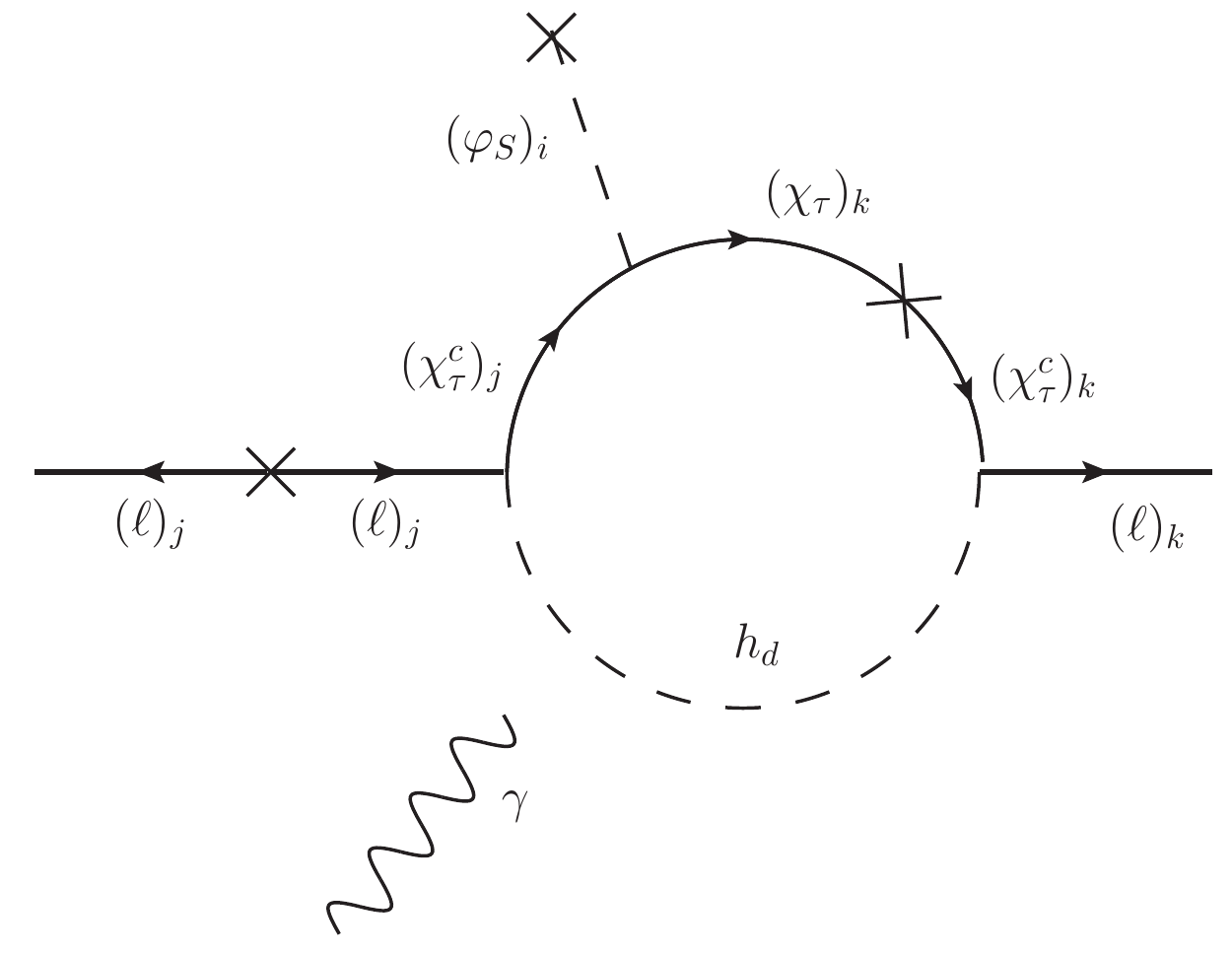}}
\caption{Diagrams for the non-canonical effective kinetic term and for the $\ell_j\rightarrow \ell_k\gamma$ process.}
\label{AM_Kinetic}
\end{figure}

\clearpage{\pagestyle{empty}\cleardoublepage}
%
%

\newpage
\bibliography{MyBiblio}
\bibliographystyle{MyStyle}

\end{document}